\documentclass[a4paper]{article}
\usepackage[ascii]{inputenc}
\usepackage{amsmath}
\usepackage{amssymb,amsfonts,textcomp}
\usepackage[T1]{fontenc}
\usepackage[english]{babel}
\usepackage{color}
\usepackage{multicol}
\usepackage{array}
\usepackage{hhline}
\usepackage{hyperref}
\hypersetup{pdftex, colorlinks=true, linkcolor=blue, citecolor=blue, filecolor=blue, urlcolor=blue, pdftitle=, pdfauthor=, pdfsubject=, pdfkeywords=}
\usepackage[pdftex]{graphicx}
\makeatletter
\newcommand\ps@Standard{
  \renewcommand\@oddhead{}
  \renewcommand\@evenhead{\@oddhead}
  \renewcommand\@oddfoot{\thepage{}}
  \renewcommand\@evenfoot{\@oddfoot}
  \renewcommand\thepage{\arabic{page}}
}
\newcommand\ps@FirstPage{
  \renewcommand\@oddhead{}
  \renewcommand\@evenhead{\@oddhead}
  \renewcommand\@oddfoot{}
  \renewcommand\@evenfoot{\@oddfoot}
  \renewcommand\thepage{\arabic{page}}
}
\makeatother
\newcounter{saveenum}
\newcommand\liststyleWWNumiii{%
\renewcommand\theenumi{\Roman{enumi}}
\renewcommand\theenumii{\Alph{enumii}}
\renewcommand\theenumiii{\Roman{enumi}.\Alph{enumii}.\arabic{enumiii}}
\renewcommand\theenumiv{\Roman{enumi}.\Alph{enumii}.\arabic{enumiii}.\arabic{enumiv}}
\renewcommand\labelenumi{\theenumi.}
\renewcommand\labelenumii{\theenumii.}
\renewcommand\labelenumiii{\theenumiii}
\renewcommand\labelenumiv{\theenumiv}
}
\newcommand\liststyleWWNumii{%
\renewcommand\theenumi{\arabic{enumi}}
\renewcommand\theenumii{\arabic{enumii}}
\renewcommand\theenumiii{\roman{enumiii}}
\renewcommand\theenumiv{\arabic{enumiv}}
\renewcommand\labelenumi{\theenumi.}
\renewcommand\labelenumii{\theenumii[3001?]}
\renewcommand\labelenumiii{\theenumiii.}
\renewcommand\labelenumiv{\theenumiv.}
}
\newcommand\liststyleWWNumi{%
\renewcommand\theenumi{\Alph{enumi}}
\renewcommand\theenumii{\Alph{enumii}}
\renewcommand\theenumiii{\Alph{enumi}.\Alph{enumii}.\arabic{enumiii}}
\renewcommand\theenumiv{\Alph{enumi}.\Alph{enumii}.\arabic{enumiii}.\arabic{enumiv}}
\renewcommand\labelenumi{\theenumi.}
\renewcommand\labelenumii{\theenumii.}
\renewcommand\labelenumiii{\theenumiii}
\renewcommand\labelenumiv{\theenumiv}
}
\newcommand\liststyleWWNumviii{%
\renewcommand\theenumi{\Alph{enumi}}
\renewcommand\theenumii{\Alph{enumii}}
\renewcommand\theenumiii{\Alph{enumi}.\Alph{enumii}.\arabic{enumiii}}
\renewcommand\theenumiv{\Alph{enumi}.\Alph{enumii}.\arabic{enumiii}.\arabic{enumiv}}
\renewcommand\labelenumi{\theenumi.}
\renewcommand\labelenumii{\theenumii.}
\renewcommand\labelenumiii{\theenumiii}
\renewcommand\labelenumiv{\theenumiv}
}
\newcommand\liststyleWWNumiv{%
\renewcommand\theenumi{\arabic{enumi}}
\renewcommand\theenumii{\arabic{enumii}}
\renewcommand\theenumiii{\roman{enumiii}}
\renewcommand\theenumiv{\arabic{enumiv}}
\renewcommand\labelenumi{\theenumi.}
\renewcommand\labelenumii{\theenumii[3001?]}
\renewcommand\labelenumiii{\theenumiii.}
\renewcommand\labelenumiv{\theenumiv.}
}
\newcommand\liststyleWWNumv{%
\renewcommand\theenumi{\arabic{enumi}}
\renewcommand\theenumii{\arabic{enumii}}
\renewcommand\theenumiii{\roman{enumiii}}
\renewcommand\theenumiv{\arabic{enumiv}}
\renewcommand\labelenumi{\theenumi.}
\renewcommand\labelenumii{\theenumii[3001?]}
\renewcommand\labelenumiii{\theenumiii.}
\renewcommand\labelenumiv{\theenumiv.}
}
\newcommand\liststyleWWNumx{%
\renewcommand\theenumi{\Alph{enumi}}
\renewcommand\theenumii{\Alph{enumii}}
\renewcommand\theenumiii{\Alph{enumi}.\Alph{enumii}.\arabic{enumiii}}
\renewcommand\theenumiv{\Alph{enumi}.\Alph{enumii}.\arabic{enumiii}.\arabic{enumiv}}
\renewcommand\labelenumi{\theenumi.}
\renewcommand\labelenumii{\theenumii.}
\renewcommand\labelenumiii{\theenumiii}
\renewcommand\labelenumiv{\theenumiv}
}
\newcommand\liststyleWWNumvi{%
\renewcommand\theenumi{\arabic{enumi}}
\renewcommand\theenumii{\arabic{enumii}}
\renewcommand\theenumiii{\roman{enumiii}}
\renewcommand\theenumiv{\arabic{enumiv}}
\renewcommand\labelenumi{\theenumi.}
\renewcommand\labelenumii{\theenumii[3001?]}
\renewcommand\labelenumiii{\theenumiii.}
\renewcommand\labelenumiv{\theenumiv.}
}
\newcommand\liststyleWWNumxi{%
\renewcommand\theenumi{\Alph{enumi}}
\renewcommand\theenumii{\Alph{enumii}}
\renewcommand\theenumiii{\Alph{enumi}.\Alph{enumii}.\arabic{enumiii}}
\renewcommand\theenumiv{\Alph{enumi}.\Alph{enumii}.\arabic{enumiii}.\arabic{enumiv}}
\renewcommand\labelenumi{\theenumi.}
\renewcommand\labelenumii{\theenumii.}
\renewcommand\labelenumiii{\theenumiii}
\renewcommand\labelenumiv{\theenumiv}
}
\newcommand\liststyleWWNumvii{%
\renewcommand\labelitemi{${\blacksquare}$}
\renewcommand\labelitemii{${\blacksquare}$}
\renewcommand\labelitemiii{${\blacklozenge}$}
\renewcommand\labelitemiv{${\bullet}$}
}
\newcommand\liststyleWWNumix{%
\renewcommand\theenumi{\arabic{enumi}}
\renewcommand\theenumii{\arabic{enumi}.\arabic{enumii}}
\renewcommand\theenumiii{\arabic{enumiii}}
\renewcommand\theenumiv{\arabic{enumi}.\arabic{enumii}.\arabic{enumiii}.\arabic{enumiv}}
\renewcommand\labelenumi{[\theenumi]}
\renewcommand\labelenumii{\theenumii)}
\renewcommand\labelenumiii{\theenumiii)}
\renewcommand\labelenumiv{\theenumiv.}
}
\title{}
\author{}
\date{}
\begin{document}
\clearpage\setcounter{page}{1}\pagestyle{Standard}
\thispagestyle{FirstPage}
{\centering
\textbf{\textcolor{black}{Toward Safe and Responsible AI Agents:}}
\par}

{\centering
\textbf{\textcolor{black}{A Three-Pillar Model for}}
\par}

{\centering
\textbf{\textcolor{black}{Transparency, Accountability, and Trustworthiness}}
\par}

\bigskip

\begin{multicols}{3}
{\centering
Edward C.\textcolor{black}{ Cheng}
\par}

{\centering
echeng04@stanford.edu
\par}

{\centering
\textcolor{black}{Jeshua} Cheng
\par}

{\centering
jeshua.cheng@inquiryon.com
\par}

{\centering
Alice Siu\newline
asiu@stanford.edu
\par}
\end{multicols}
[Warning: Draw object ignored]

\bigskip

\textbf{\textit{Abstract}}\textit{ -- This paper presents a conceptual and operational framework for developing and
operating safe and trustworthy AI agents based on a Three-Pillar Model grounded in transparency, accountability, and
trustworthiness. Building on prior work in Human-in-the-Loop systems, reinforcement learning, and collaborative AI, the
framework defines an evolutionary path toward autonomous agents that balances increasing automation with appropriate
human oversight. The paper argues that safe agent autonomy must be achieved through progressive validation, analogous
to the staged development of autonomous driving, rather than through immediate full automation. Transparency and
accountability are identified as foundational requirements for establishing user trust and for mitigating known risks
in generative AI systems, including hallucinations, data bias, and goal misalignment, such as the inversion problem.
The paper further describes three ongoing work streams supporting this framework: public deliberation on AI agents
conducted by the Stanford Deliberative Democracy Lab, cross-industry collaboration through the Safe AI Agent
Consortium, and the development of open tooling for an agent operating environment aligned with the Three-Pillar Model.
Together, these contributions provide both conceptual clarity and practical guidance for enabling the responsible
evolution of AI agents that operate transparently, remain aligned with human values, and sustain societal trust.}

\bigskip

\textbf{\textit{Keywords--- Generative AI, AI Agent, Human-in-the-Loop, HITL, RLHF, Responsible AI, Trustworthy AI}}

{\centering [Warning: Draw object ignored]\par}

\bigskip

\liststyleWWNumiii
\begin{enumerate}
\item {\scshape\color{black}
Introduction}
\end{enumerate}
\textcolor{black}{The emergence of AI agents marks a new phase in the evolution of generative AI. While traditional
chatbots focus on generating text-based responses, AI agents extend this capability into real-world action. These
systems can execute tasks, reason over goals, and make decisions on behalf of humans. This shift from text generation
to autonomous task execution holds the key to unlocking the economic and practical value of generative AI. Yet, as
these systems gain autonomy and agency, the risks of error, bias, and misalignment also multiply. When AI agents make
consequential real-life decisions, such as transferring funds, filing drug prescriptions, drafting contracts, or
guiding robotic actions, their mistakes may lead to financial losses, privacy breaches, or even physical harms. These
errors may arise from training biases, lack of situational context, hallucinated reasoning, or misalignment between
user intent and model objectives. Consequently, the field is confronted with an urgent challenge: how to ensure safe,
transparent, and accountable AI agents that enhance productivity without compromising accuracy, trust or human values.}

\textcolor{black}{A growing body of recent research has emerged to address this challenge, focusing on the
Human-in-the-Loop (HITL) paradigm and its extensions as means to govern, calibrate, and align AI agent behavior. These
works explore how human expertise, oversight, and ethical grounding can be woven into the AI learning and action loop
to produce systems that are both impactful and controllable. Collectively, they represent a growing consensus that
human-AI collaboration, rather than full automation, is the most promising pathway toward efficient, effective, and
safe AI agents that will result in higher productivity gain [1].}

\textcolor{black}{To organize this literature, we can group the representative surveys into three major thematic
clusters that trace the conceptual evolution of safe AI agent design:}

\liststyleWWNumii
\begin{enumerate}
\item \textcolor{black}{Foundational theories of human-in-the-loop AI and machine learning.}
\item \textcolor{black}{Operational frameworks and platforms for human-AI collaboration.}
\item \textcolor{black}{Emerging approaches for uncertainty alignment and human-governed AI agents.}
\end{enumerate}
\liststyleWWNumi
\begin{enumerate}
\item {\bfseries\itshape\color{black}
Foundational Theories of Human-in-the-Loop AI}
\end{enumerate}
\textcolor{black}{Early research established the theoretical and ethical foundations for integrating humans into the AI
lifecycle. Zanzotto (2019) proposed Human-in-the-loop Artificial Intelligence (HitAI) as both a moral and structural
correction to the unregulated growth of autonomous AI [2]. He argued that humans are not mere annotators but the
original ``knowledge producers'' whose insights underpin AI performance and thus must remain central to both credit and
control. Wu et al. (2022) expanded this notion through a systematic survey of HITL for machine learning, framing it as
a data-centric methodology that unites human cognition with computational scalability. They demonstrated that effective
human involvement improves labeling efficiency, interpretability, and robustness, forming the foundation for iterative
feedback loops in model development [3].}

\textcolor{black}{Building on these theoretical bases, Mosqueira-Rey et al. (2023) presented a unifying taxonomy of
Human-in-the-Loop Machine Learning (HITL-ML) paradigms [4]. They identified key interaction modes, which include Active
Learning, Interactive ML, Machine Teaching, Curriculum Learning, and Explainable AI. They revealed that human-AI
relationships exist along a continuum of control: from machine-driven query optimization to human-driven knowledge
transfer and interpretation. These early frameworks collectively redefined HITL as not simply supervision, but shared
agency between human reasoning and machine inference, setting the epistemic groundwork for subsequent advances in
safety and transparency.}

\textcolor{black}{Extending the HITL perspective beyond technical design, recent studies from MIT Sloan introduced a
management-oriented framework known as AI Alignment. This paradigm emphasizes that model accuracy, reliability in
real-world contexts, and stakeholder relevance must be achieved through continuous human engagement. It reframes human
involvement not only as a safeguard but also as a means for organizations to learn and adapt as they deploy AI.
Grounded in empirical case studies, this framework shows that practices such as expert feedback and stakeholder
participation are essential for building safe, context-aware AI systems [5]. A complementary MIT Sloan study found that
asking critical safety questions early in the AI development process helps prevent systemic errors and security
vulnerabilities, further reinforcing the importance of proactive human oversight [6].}

\liststyleWWNumi
\setcounter{saveenum}{\value{enumi}}
\begin{enumerate}
\setcounter{enumi}{\value{saveenum}}
\item {\bfseries\itshape\color{black}
Operational Frameworks for Safe and Collaborative AI Agents}
\end{enumerate}
\textcolor{black}{As Human-in-the-Loop principles matured, a second wave of research shifted toward practical frameworks
and system architectures that enable effective human-AI collaboration in real-world, embodied environments. Bellos and
Siskind (2025) exemplify this transition by introducing a structured evaluation framework, a multimodal dataset, and an
augmented-reality (AR) AI agent designed to guide humans through complex physical tasks such as culinary cooking and
battlefield medicine. Their empirical studies demonstrate that interactive, context-aware guidance significantly
improves task success rates, reduces procedural errors, and enhances user experience. Importantly, their results also
show that exposure to AI-assisted guidance leads to measurable improvements in subsequent unassisted task performance,
indicating that AI agents can support not only immediate task completion but also longer-term human skill acquisition.
These findings position AI agents as collaborative partners that augment human capability rather than as purely
automated systems [7].}

\textcolor{black}{In parallel, Mozannar et al. (2025) introduced
}\textit{\textcolor{black}{Magentic-UI}}\textcolor{black}{, an open-source user-interface platform for
human-in-the-loop agentic systems. Built on Microsoft's }\textit{\textcolor{black}{Magentic-One}}\textcolor{black}{
framework, it enables users to co-plan, co-execute, approve, and verify AI actions in complex digital tasks such as
coding and document handling [8]. The platform embeds human oversight through structured, repeatable mechanisms. It
supports co-planning, co-tasking, action approval, and answer verification, establishing a controlled environment for
studying trust calibration, safety, and usability in AI agents. Together, these efforts move the field from abstract
advocacy to practical system engineering, demonstrating that safety and transparency can be designed into agent
interfaces, workflows, and orchestration protocols.}

\liststyleWWNumi
\setcounter{saveenum}{\value{enumi}}
\begin{enumerate}
\setcounter{enumi}{\value{saveenum}}
\item {\bfseries\itshape\color{black}
Emerging Approaches for Uncertainty-Aware and Human-Governed AI Agents}
\end{enumerate}
\textcolor{black}{Recent work has deepened the mathematical and procedural foundations of safety and alignment. Retzlaff
et al. (2024) surveyed the domain of Human-in-the-Loop Reinforcement Learning (HITL-RL), arguing that reinforcement
learning (RL) inherently depends on human feedback and should be understood as a HITL paradigm. Their work outlined
design requirements such as feedback quality, trust calibration, and explainability for moving from human-guided to
human-governed learning [9]. Complementing this, Ren et al. (2023) proposed the KNOWNO (``Know When You Don't Know'')
framework for LLM-driven robotic planners to identify critical moments that require human involvement. By employing
conformal prediction to quantify uncertainty, KNOWNO enables robots to detect when their confidence falls below a
safety threshold and proactively request human input to ensure safe and reliable task execution [10]. This model of
uncertainty alignment provides formal statistical guarantees on task success while minimizing unnecessary human
intervention. This work represents a crucial step toward self-aware, help-seeking agents.}

\textcolor{black}{At a broader institutional level, research from Harvard University has expanded the discussion of AI
safety to include ethics, governance, and societal accountability. Allen et al. (2024) proposed a democratic model of
power-sharing liberalism, emphasizing human flourishing, shared authority, and institutional accountability. They
argued that AI governance must move beyond risk management to actively promote public goods, equality, and autonomy
through inclusive participation and transparent oversight. Their framework identifies six core governance tasks:
mitigating harm, managing emergent capabilities, preventing misuse, advancing public benefit, building human capital,
and strengthening democratic capacity [11]. Complementing this perspective, Barroso and Mello (2024) examined AI as
both a revolutionary and perilous force shaping humanity's future, calling for a global governance framework grounded
in human dignity, transparency, accountability, and democratic oversight [12]. Together, these contributions frame AI
not as a force to restrain but as a catalyst for renewing democracy and reinforcing collective well-being.}

\textcolor{black}{Finally, Natarajan et al. (2025) reframed the entire discussion through the concept of AI-in-the-Loop
(AI2L). Their analysis reveals that many systems labeled as HITL should be considered as AI2L, where humans, not AI,
remain the decision-makers. They argue that this distinction is critical for designing systems that emphasize
collaboration over automation, human impact over algorithmic efficiency, and co-adaptive intelligence over substitution
[13]. This reorientation marks a philosophical inflection point: moving from human-assisted AI to AI-assisted
humanity.}

\liststyleWWNumi
\setcounter{saveenum}{\value{enumi}}
\begin{enumerate}
\setcounter{enumi}{\value{saveenum}}
\item {\bfseries\itshape\color{black}
Toward a Framework for Safe, Transparent AI Agents}
\end{enumerate}
\textcolor{black}{Across these studies, a clear trajectory emerges. The field has progressed from recognizing the
ethical necessity of human oversight, to engineering collaborative systems, and to developing experimentally grounded
mechanisms for uncertainty and governance. Collectively, these efforts affirm that the challenge of AI agent safety,
transparency, and alignment is both urgent and tractable. Embedding humans as teachers, collaborators, and governors
within the AI lifecycle consistently improves reliability and trustworthiness, yet fragmentation persists across
methodologies and evaluation metrics.}

\textcolor{black}{This paper advances the next step to synthesize these developments into a unified conceptual framework
and a set of guiding principles that integrate HITL, AI2L, uncertainty alignment, and human-governed learning into a
progressively improving autonomous environment. Together, these foundations define an operational setting for a new
generation of AI agents that are transparent by design, collaborative by nature, and accountable in operation, with the
explicit goal of enabling increasing level of autonomy in a safe, controlled, and trustworthy manner.}

\liststyleWWNumiii
\begin{enumerate}
\item {\scshape\color{black}
The Evolution Path Towards Autonomous Agents}
\end{enumerate}
\textcolor{black}{The vision of achieving fully autonomous AI agents represents one of the most ambitious goals in
artificial intelligence. However, this vision cannot be realized in a single leap. It must evolve through progressive
stages of validation and oversight, where human involvement is reduced only as confidence in the system's performance
and alignment grows through proven safety, reliability, and accountability. This evolutionary approach has clear
precedents in other industries, particularly in the development of autonomous driving.}

\liststyleWWNumviii
\begin{enumerate}
\item {\bfseries\itshape\color{black}
Lessons from Autonomous Driving}
\end{enumerate}
\textcolor{black}{The field of autonomous driving provides an instructive example of how automation can evolve
responsibly. Early driver-assist systems such as adaptive cruise control and lane-keeping support were designed to
assist rather than replace human judgment. These systems required the driver to maintain foot on the pedal, hands on
the wheel, and eyes on the road at all times. As perception models, control algorithms, and sensor fusion technologies
advanced, vehicles began to handle more complex scenarios independently, such as automatic parking and highway lane
changes. At this stage, the human driver could briefly disengage from active control but still had to monitor the road
and be prepared to intervene if necessary.}

 \includegraphics[width=6.2681in,height=1.1728in]{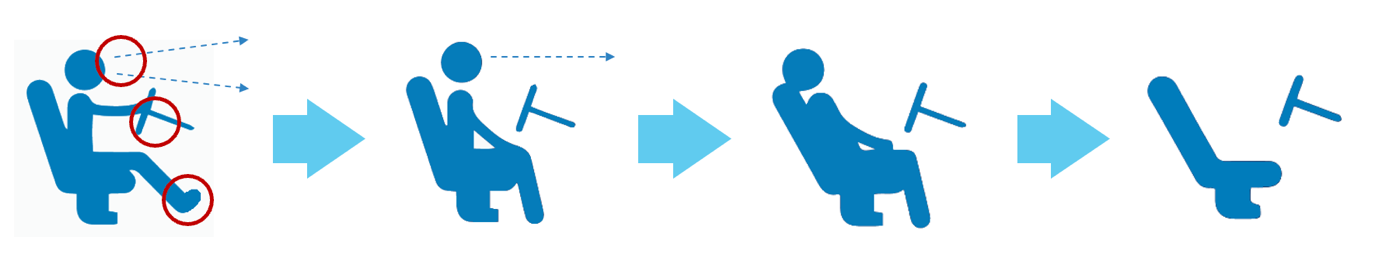} 

{\centering
\textbf{\textcolor{black}{Figure 1: Full Autonomous Driving was a Gradual Evolving Process}}
\par}

\textcolor{black}{This gradual and transparent evolution allowed engineers to identify edge cases, improve algorithms,
and refine user interfaces based on real-world feedback. Most importantly, it allowed trust of the society to grow
incrementally. Each technological improvement was accompanied by clearer communication about the system's limitations
and capabilities. Drivers learned when to rely on the system and when to take over. Through testing, validation, and
iterative learning, both the technology and its human users matured on autonomous driving together. Only through this
patient process did the industry approach }\textit{\textcolor{black}{Level 4 and Level 5 autonomy}}\textcolor{black}{,
where vehicles can operate without human intervention in most or all conditions [12, 13]. The success of this journey
lies not only in technological innovation but also in earning human trust through transparency, communication of system
limits, and clear accountability.}

\liststyleWWNumviii
\setcounter{saveenum}{\value{enumi}}
\begin{enumerate}
\setcounter{enumi}{\value{saveenum}}
\item {\bfseries\itshape\color{black}
Parallels in AI Agents Development}
\end{enumerate}
\textcolor{black}{A similar path must be followed in the evolution of autonomous AI agents. These systems act on behalf
of humans in both digital and physical environments, making decisions that can have significant consequences. Like
early autonomous vehicles required drivers to remain attentive, current AI agents still depend on Human-in-the-Loop
(HITL) oversight to ensure that their actions align with human intent. Human involvement serves as both a safeguard and
a source of learning, helping the system adapt responsibly. As discussed in the earlier introduction, research by Wu
(2022), Mosqueira-Rey (2023), and Retzlaff (2024) consistently shows that HITL systems improve interpretability,
accountability, and model reliability [3, 4, 9]. Rather than viewing human oversight as an administrative overhead, it
should be recognized as a critical step in the learning and governance process that helps agents mature progressively.}

\liststyleWWNumviii
\setcounter{saveenum}{\value{enumi}}
\begin{enumerate}
\setcounter{enumi}{\value{saveenum}}
\item {\bfseries\itshape\color{black}
HITL as a Mechanism for Trust and Safety}
\end{enumerate}
\textcolor{black}{Human oversight is particularly essential during the intermediate stages of agent development and
deployment. At this point, agents are capable of complex reasoning but still lack the contextual, ethical, and
situational awareness required for independent operation [16]. Well-designed HITL mechanisms allow humans to validate
outputs, correct errors, and prevent harm caused by hallucinations, data biases, or incorrect assumptions. This
feedback loop not only safeguards users but also enables the system to learn and improve over time. As the system
demonstrates consistent accuracy and reliability, the level of human intervention can be reduced. However, this
reduction must be based on measurable improvements, not assumption.}

\textcolor{black}{The importance of this gradual approach becomes even more evident in trust-sensitive domains such as
finance, human resources, healthcare, legal, and areas that require regulatory compliance. Human oversight ensures
share responsibility between humans and AI, maintaining compliance with both legal standard and societal expectations.
Just as self-driving systems underwent years of supervised testing before being trusted on public roads, autonomous AI
agents must demonstrate reliability before operating independently in high-stakes environments. Yet, their journey to
full automation will likely unfold more rapidly, driven by the accelerating pace of AI research and development.}

\liststyleWWNumviii
\setcounter{saveenum}{\value{enumi}}
\begin{enumerate}
\setcounter{enumi}{\value{saveenum}}
\item {\bfseries\itshape\color{black}
A Collaborative Path Toward Full Autonomy}
\end{enumerate}
\textcolor{black}{The journey toward fully autonomous agents is both a technological and social process. Technological
progress enables higher levels of independence, while social acceptance depends on observable safety and
accountability. Research from Bellos (2025) and Mozannar (2025) has shown that when humans and AI collaborate
effectively, the result is higher task success rates, improved trust, and greater user confidence [7, 8]. Collaboration
thus provides a bridge between current assisted systems and the future of full autonomy.}

\textcolor{black}{This process can be viewed as four evolutionary stages of AI agency:}

\liststyleWWNumiv
\begin{enumerate}
\item \textbf{\textcolor{black}{Assisted Agents: }}\textcolor{black}{Humans make decisions while AI supports them
through recommendations and reasoning.}
\item \textbf{\textcolor{black}{Collaborative Agents:}}\textcolor{black}{ Humans and AI share responsibility in
decision-making and task execution, combining human contextual understanding with AI computational precision and
scalability. Human participation remains essential within the agentic workflow, as it enriches situational and semantic
context, ensuring that AI agents produce responses and actions that are relevant, accurate, and aligned with user
intent and real-world constraints [16].}
\item \textbf{\textcolor{black}{Supervised Autonomy: }}\textcolor{black}{AI operates independently in constrained
environments while remaining accountable through human review.}
\item \textbf{\textcolor{black}{Full Autonomy with Human Governance:}}\textcolor{black}{ AI functions independently
within transparent, auditable frameworks that preserve human oversight at the policy level.}
\end{enumerate}
\textcolor{black}{Advancement through these stages must be validated by evidence of safety, predictability, and
alignment with human intent. This progressive process reflects the same progression that made autonomous driving
successful. Skipping these steps would risk premature deployment and loss of confidence, which could set back both
innovation and adoption.}

\liststyleWWNumviii
\begin{enumerate}
\item {\bfseries\itshape\color{black}
Toward Trustworthy Autonomy}
\end{enumerate}
\textcolor{black}{True autonomy cannot be declared by design; it must be demonstrated through experience and data. Each
stage of progress should confirm that the agent can act responsibly and transparently within defined boundaries. By
embedding Human-in-the-Loop principles throughout the development process ensures that autonomy and trust grow in
tandem. As seen in autonomous driving, confidence arises from steady progress and accountable design. While AI agents
may reach maturity more quickly due to faster digital feedback loops and lower physical risks, their path to autonomy
must still be guided by the same principles of transparency, validation, and ethical oversight.}

\liststyleWWNumiii
\begin{enumerate}
\item {\scshape\color{black}
A Three-Pillar Model for a Safe AI-Agent Operating Environment}
\end{enumerate}
\textcolor{black}{In the previous sections, we demonstrated that as AI systems evolve from passive chatbots to fully
autonomous agents capable of acting on behalf of humans, the potential for both benefit and harm expands dramatically.
In addition, as AI agents evolve to become increasingly independent of humans, their autonomy must emerge through a
gradual, trust-building process in which human oversight and collaboration remain essential until AI systems
demonstrate consistent reliability and alignment.}

\textcolor{black}{Building on these foundations, this section proposes that to enable this evolutionary process to
unfold safely and productively, AI agents must operate within a structured environment designed to support growth,
supervision, and accountability. Without such an environment, autonomous evolution would occur in an uncontrolled
manner, exposing organizations and individuals to unacceptable risks.}

\textcolor{black}{To address this need, we propose a Three-Pillar Model (3PM) to support a safe AI-agent operating
environment. This model defines the fundamental principles and environmental conditions required to develop, deploy,
and operate safe autonomous agents while maintaining a balance between automation and human collaboration. The three
pillars are:}

\liststyleWWNumv
\begin{enumerate}
\item \textbf{\textcolor{black}{Transparency of AI Agents}}\textcolor{black}{ ensures visibility into how agents operate
across their life cycles.}
\item \textbf{\textcolor{black}{Accountability in Decision-Making}}\textcolor{black}{ provides mechanisms to attribute
and explain decisions made by both humans and AI.}
\item \textbf{\textcolor{black}{Trustworthiness through Human-AI Collaboration}}\textcolor{black}{ establishes
confidence in agentic systems through well-timed human oversight and fallback safeguards.}
\end{enumerate}
\textcolor{black}{Together, these pillars create the foundation for a safe and productive ecosystem where AI agents and
humans can share responsibilities and co-evolve toward higher levels of autonomy. They support the long-term goal of
achieving responsible, human-aligned AI while ensuring that enterprises can realize measurable return on investment
through efficient, reliable, and trustworthy automation.}

\liststyleWWNumx
\begin{enumerate}
\item {\bfseries\itshape\color{black}
Pillar One: Transparency and Building Trust with AI}
\end{enumerate}
\textcolor{black}{Transparency provides the visibility necessary for humans to understand, monitor, guide, and audit
agent behavior. It allows operators to know how the agent works, what it is doing, and why it acts in a particular way.
This visibility is critical during the evolutionary path described earlier, because it enables humans to supervise and
calibrate the agent's performance as autonomy increases.}

\textcolor{black}{Every agent instance passes through a lifecycle consisting of three stages: initiation, active
operation, and completion or termination. Transparency must exist throughout each stage to make the process
comprehensible and auditable.}

{\centering  \includegraphics[width=4.9846in,height=2.3327in]{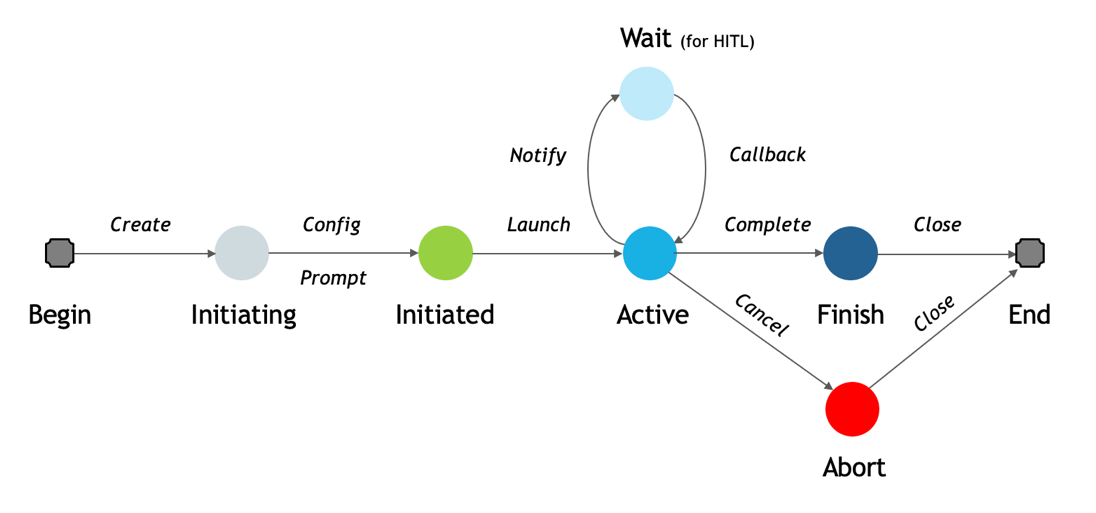} \par}
{\centering
\textbf{\textcolor{black}{Figure 2 Agent State Transition Diagram Within the 3-Pillar Model}}
\par}

\textbf{\textcolor{black}{Initiation State.}}\textcolor{black}{ During initiation, a human defines the scope, context,
and objectives of the agent's work. This stage establishes the foundation for safe collaboration. For example, a
Research Agent tasked with supporting a product's go-to-market strategy must receive a clearly defined configuration
that includes market segments, data sources, and success criteria. By setting these parameters, the human ensures that
the agent's goals are properly aligned with organizational objectives and ethical standards. This stage also serves as
a point of human control, where configurations, role definitions, and constraints can be verified before the agent
begins operation.}

\textbf{\textcolor{black}{Active State. }}\textcolor{black}{Once launched, the agent enters its active state, where it
performs the actions for which it was designed. For instance, a Research Agent may conduct web searches and synthesize
findings. Likewise, a Payment Agent may initiate payment transactions. A Collection Letter Agent may draft personalized
communications based on debtor information and credit conditions. During this phase, activity recording and
observability become essential. The environment must automatically generate activity journals that record the agent's
decisions, interactions, and results.}

\textcolor{black}{These logs enable oversight and provide a transparent record for post-task evaluation. Moreover,
during this phase, the Human-in-the-Loop (HITL) mechanism plays an important role. When the agent encounters
uncertainty or ambiguity, it may consult a human collaborator for guidance. Depending on task complexity and risk
level, human involvement can vary from direct supervision to collaborative decision-making to minimal observation.
Transparency allows both sides to know when and why such handoffs occur.}

\textbf{\textcolor{black}{Abort State. }}\textcolor{black}{Both human operators and authorized AI subsystems should have
the ability to abort or suspend an active agent when necessary. Abort events may occur if the agent cannot fulfill its
mission due to missing resources, time constraints, or safety violations. The authority to abort should follow clearly
defined governance rules, reflecting the contractual and regulatory conditions under which the agent operates.}

\textbf{\textcolor{black}{Finish State. }}\textcolor{black}{When an agent finishes or terminates its task, it should
produce a clear output along with a record of its entire operation. Transparency requires three complementary forms of
documentation:}

\liststyleWWNumvi
\begin{enumerate}
\item \textbf{\textcolor{black}{State transition records: }}\textcolor{black}{Marking changes from initiation to
finish.}
\item \textbf{\textcolor{black}{Work progress records: }}\textcolor{black}{Showing the detailed actions taken by the
agent.}
\item \textbf{\textcolor{black}{HITL records: }}\textcolor{black}{capturing every human--AI interaction and decision.}
\end{enumerate}
\textcolor{black}{These records serve as the backbone of transparency within the agent operating environment. They allow
developers, regulators, and users to reconstruct events, assess system performance, and identify opportunities for
improvement. Without sufficient transparency, human collaborators cannot effectively supervise agent behavior, learn
from outcomes, or develop trust in autonomous agent systems. While these three record types are not exhaustive, they
represent the minimum information required to achieve acceptable transparency. In practice, the agent system may also
maintain additional journals, such as system logs, user feedback logs, performance metrics, and other operational
traces, to further support monitoring, analysis, and continuous improvement.}

\liststyleWWNumx
\begin{enumerate}
\item {\bfseries\itshape\color{black}
Pillar Two: Accountability and Responsibility}
\end{enumerate}
\textcolor{black}{While transparency answers what happened, accountability answers why it happened and who is
responsible. In the previous section on the evolutionary path, we emphasized that autonomy must be earned gradually.
Accountability provides the ethical and operational framework that makes this process safe. As AI agents gain more
independence, the environment must ensure that each decision, whether made by a human or AI, is traceable to its source
and understandable and explainable in context.}

\textcolor{black}{Achieving accountability requires comprehensive decision journaling that records not only the outcomes
but also the reasoning and contextual factors behind each choice. This is closely related to the principle of
explainability in AI. Agents must be able to provide, upon request, the rationale for their decisions, including the
data sources consulted, the constraints considered, and the degree of confidence associated with their outputs.}

\textcolor{black}{A practical example illustrates this need. Suppose an automated food-ordering agent failed to account
for a customer's allergy to wheat or soy, resulting in a serious medical incident. In such a case, assigning
responsibility requires a clear understanding of each participant's role in the agentic workflow. Was the customer's
input ambiguous? Did a human worker at the restaurant fail to verify the order details during preparation? Did the AI
agent miscommunicate the constraints? Or did the underlying language model generate an inaccurate summary of the order
that omitted critical information? Without explicit records of each decision and the reasoning behind it, no clear
accountability can be established or assigned.}

\textcolor{black}{Accountability serves both corrective and developmental purposes. From a legal or regulatory
perspective, it ensures that organizations can assign responsibility when things go wrong. From a technical
perspective, it enables learning and continuous improvement. By identifying which part of the agentic workflow led to
an undesirable outcome, AI and engineers can make targeted improvements to prevent recurrence. Accountability thus
becomes the engine of continuous improvement within the agent ecosystem, reinforcing the learning loop necessary for
safe autonomy and growing trust.}

\liststyleWWNumx
\setcounter{saveenum}{\value{enumi}}
\begin{enumerate}
\setcounter{enumi}{\value{saveenum}}
\item {\bfseries\itshape\color{black}
Pillar Three: Trustworthiness \& Human-in-the-Loop}
\end{enumerate}
\textcolor{black}{The third pillar, trustworthiness, unites and build on top of the previous two. Transparency makes
operations visible, accountability clarifies responsibility, and trustworthiness converts these attributes into
confidence and willingness to rely on autonomous systems.}

\textcolor{black}{As discussed in the evolutionary path section, human trust is not granted by design but earned through
consistent, observable, and reliable performance. During the early phases of adoption, enterprises and end users will
trust AI agents only if they can see clear boundaries of control and know that humans can intervene when necessary.
Therefore, the operating environment must include mechanisms to specify risk thresholds and escalation rules that
determine when human oversight is required.}

\textcolor{black}{For example, in domains such as finance or healthcare, high-risk actions such as large transactions or
clinical recommendations should automatically trigger human review. These checkpoints form structured Human-in-the-Loop
interventions that ensure oversight at critical moments. Conversely, in high-volume, low-risk tasks, AI may operate
independently for greater efficiency. Over time, as the system demonstrates reliability, the frequency of human
interventions can be gradually reduced, following the same incremental trust-building logic that was illustrated in the
autonomous driving analogy. However, any decision to increase the level of autonomy must be explicitly approved by a
human authority and clearly documented. In addition, periodic spot checks should be conducted to verify safety and
correctness, even after incremental advances in autonomous decision-making have been introduced.}

\textcolor{black}{Trustworthiness also recognizes that in some contexts, AI can be more dependable than humans. Machines
do not suffer from fatigue, emotional fluctuation, or inconsistency, and in repetitive or data-intensive tasks, AI may
exhibit higher reliability than human operators. Accordingly, a trustworthy operating environment must support mutual
confidence. Humans must trust AI agents to function within clearly defined safety boundaries, while AI systems must be
designed to rely on validated human inputs and to defer judgment appropriately when required. The objective is not
blind reliance but calibrated trust, grounded in empirical performance evidence and shared accountability. To support
this calibration, every decision and every change must be properly recorded and remain auditable.}

\textcolor{black}{Finally, trustworthiness ensures that when failures occur, they do not propagate unchecked. The
environment must include robust fallback and recovery mechanisms that detect anomalies based on historical patterns,
suspend automated actions, and transfer control to human operators before harm occurs. These safety measures ensure
that risk remains manageable in very large-scale deployments with thousands of concurrently operating agents, even as
autonomy levels continue to increase.}

\liststyleWWNumx
\setcounter{saveenum}{\value{enumi}}
\begin{enumerate}
\setcounter{enumi}{\value{saveenum}}
\item {\bfseries\itshape\color{black}
Integrating the Three Pillars in the Evolutionary Process}
\end{enumerate}
\textcolor{black}{The Three-Pillar Model is not a theoretical abstraction but a practical extension of the evolutionary
approach described earlier. As agents progress from Assisted to Collaborative, to Supervised Autonomy, and ultimately
to Full Autonomy under Human Governance, the balance among the three pillars must evolve in parallel with each
successive stage of autonomy.}

\textcolor{black}{In early stages, transparency plays the dominant role, ensuring that every action is observable,
explainable, and auditable. As systems progress into collaborative stages, accountability becomes increasingly
important because humans and AI share responsibility for decisions and outcomes. In the later stages, once agents have
demonstrated consistent reliability and alignment, trustworthiness becomes the decisive factor that enables increasing
levels of autonomy. Importantly, companies and users will always retain the ability to determine the degree of autonomy
they are comfortable and willing to grant to different agents operating in their environments. This flexibility allows
organizations to balance efficiency with risk tolerance, enabling a gradual and confident transition toward greater
autonomy while maintaining control and trust throughout the process.}

\textcolor{black}{These pillars together form a feedback ecosystem in which humans and AI learn from each other.
Transparency provides data for accountability. Accountability identifies what needs improvement. Trustworthiness
motivates greater delegation of control. Through this cycle, autonomy grows safely and progressively.}

\textcolor{black}{In conclusion, the 3PM for agent creation, deployment, and operation establishes the essential
conditions for safe evolution toward autonomous agents. It ensures that the journey from collaboration to independence
occurs within a structure that is observable, responsible, and trustworthy. Only through such an environment can
enterprises accelerate adoption, build user confidence, and achieve the full potential of AI agents while preserving
human values and safety.}

\liststyleWWNumiii
\begin{enumerate}
\item {\scshape\color{black}
A Sample Use Case: Group Email Agent}
\end{enumerate}
\textcolor{black}{To illustrate the application of the Three-Pillar Model within a practical context, we consider a
Group Email Agent operating in an enterprise-grade agentic environment. This use case demonstrates how transparency,
accountability, and trustworthiness jointly ensure safe and effective collaboration between humans and AI.}

\textcolor{black}{A Group Email Agent is a common and valuable application for enterprises that need to compose, review,
and distribute communications to internal employees, customers, or business partners. Such messages can include policy
updates, marketing announcements, product release communications, event invitations, or crisis management
notifications. Because of their wide impact, group emails typically require coordination among multiple stakeholders,
including representatives from the business unit, marketing and communications teams, legal and compliance departments,
and senior management. These participants contribute to drafting, editing, verifying, and approving both the message
content and the list of recipients. The Group Email Agent acts as an author, a coordinator, and executor, automating
repetitive tasks while preserving human oversight where contextual understanding and judgment are critical.}

\textcolor{black}{Figure 3 displays the agent activity records captured by the system throughout the lifecycle of a
Group Email Agent instance. These records include state transitions, detailed task progress, and Human-in-the-Loop
interactions, illustrating how the operating environment maintains continuous transparency and traceability from
initiation to completion.}

{\centering  \includegraphics[width=5.3992in,height=5.2043in]{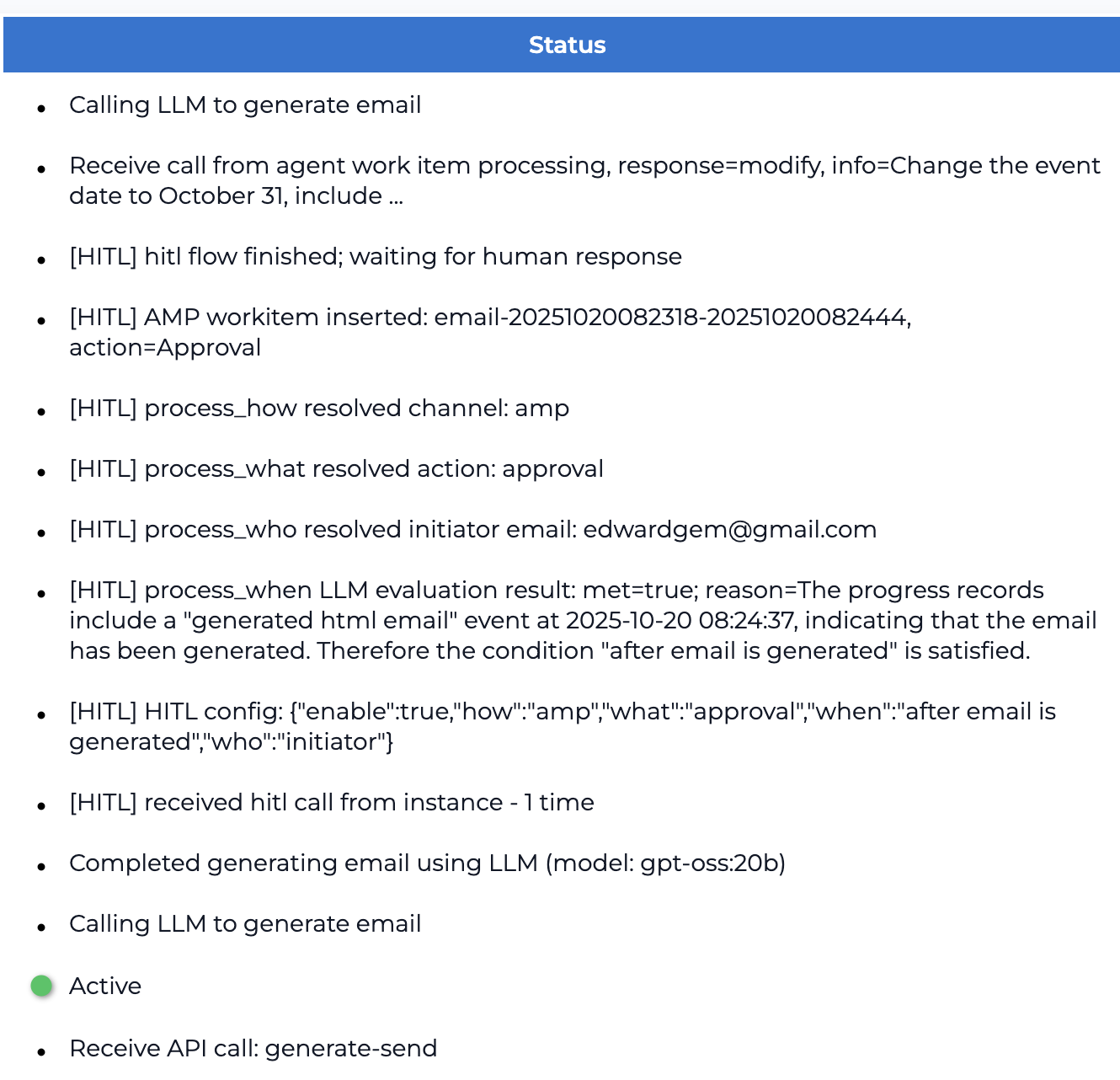} \par}
{\centering
\textbf{\textcolor{black}{Figure 3: Agent Activity Records Captured While Running a Group Email Agent}}
\par}

\bigskip

\textcolor{black}{Figure 4 demonstrates a user interface (UI) portal that enables an authorized human participant to
provide contextual inputs, review agent activities, and intervene when necessary. This interface supports
Human-in-the-Loop collaboration by allowing users to configure, guide, or correct the agent's actions in real time,
ensuring that human oversight remains an integral part of the agent's operational workflow.}

\bigskip

{\centering  \includegraphics[width=6.2681in,height=1.9516in]{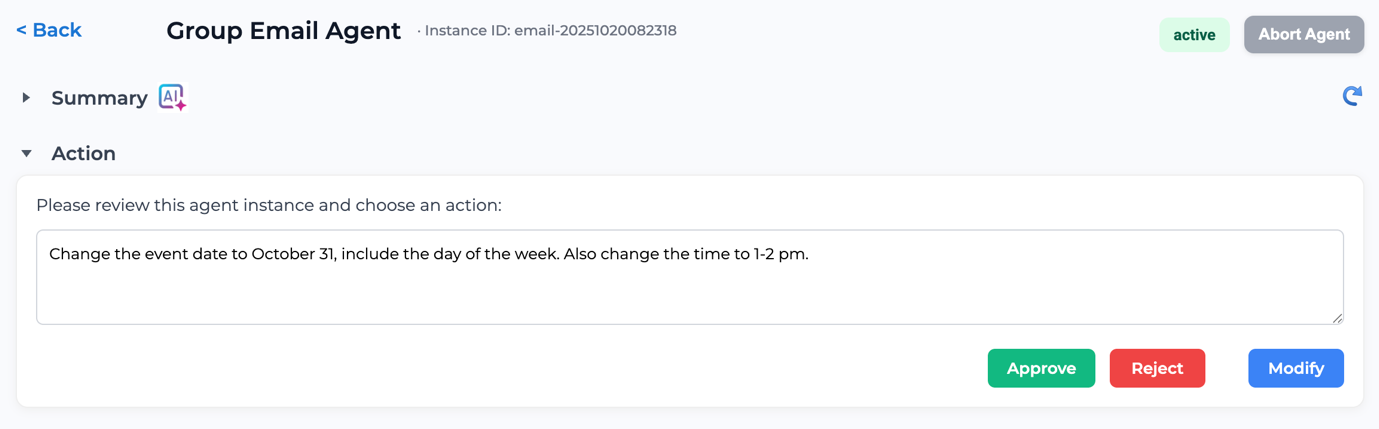} \par}
{\centering
\textbf{\textcolor{black}{Figure 4: UI for Obtaining Human-in-the-Loop Inputs During an Agentic Workflow}}
\par}

\bigskip

\textcolor{black}{Figure 5 illustrates how users can interact with the large language model (LLM) to discover, query,
and inspect the progress of agents operating within the environment. Through conversational interfaces, users can
retrieve explanations, review activity logs, and monitor task completion status. This interactive transparency fosters
mutual understanding and trust between humans and AI agents, allowing confidence to grow naturally as agents
demonstrate reliability and accountability over time.}

{\centering  \includegraphics[width=6.2681in,height=3.6693in]{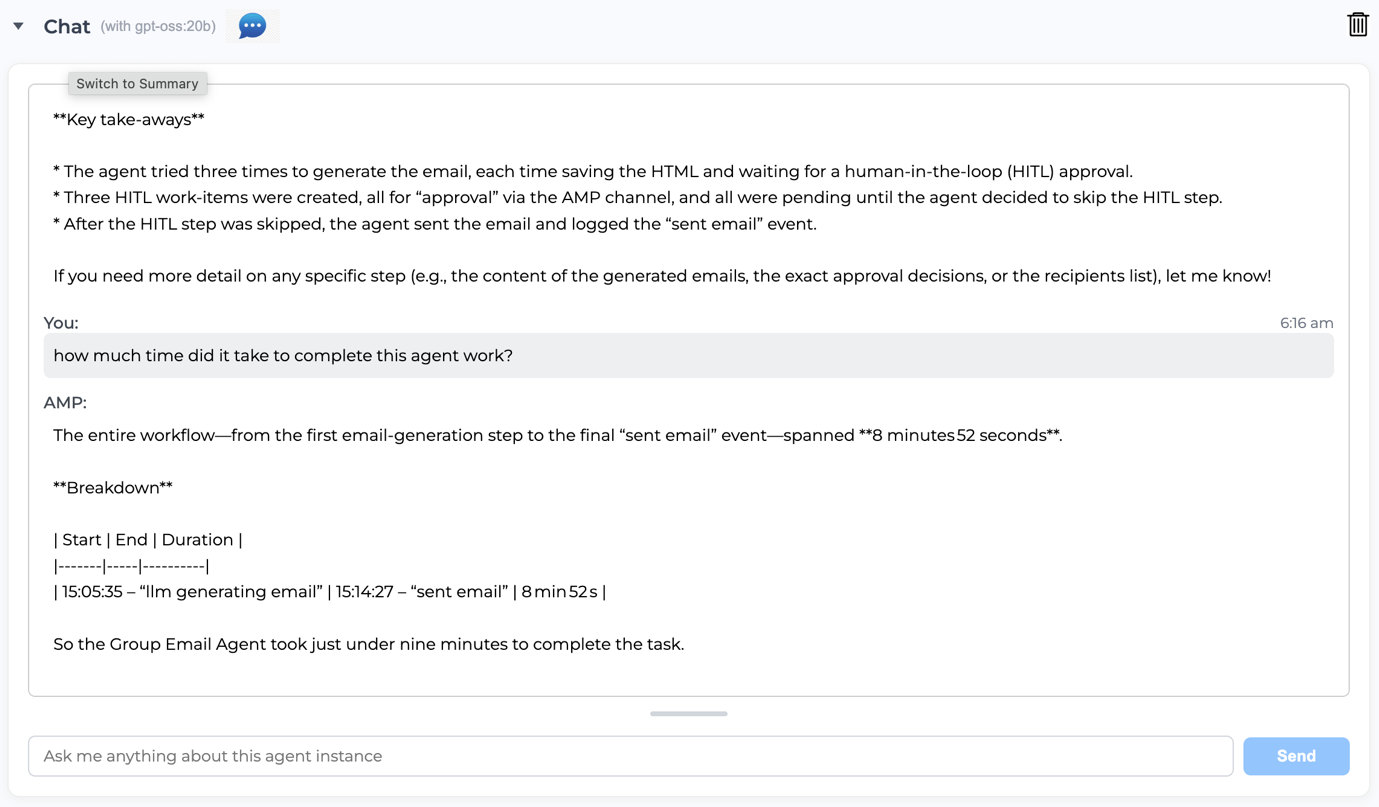} \par}
{\centering
\textbf{\textcolor{black}{Figure 5: Transparency Enables Human-AI Collaboration with Trustworthiness in Agent
Operation}}
\par}

\bigskip

\textcolor{black}{By systematically capturing and recording agent activities, the operating environment enables a high
degree of transparency that supports comprehensive analytics on both agent behavior and Human-in-the-Loop interactions.
This transparency makes it possible to surface aggregated insights through a dashboard component, which serves as a
central interface for monitoring, managing, and improving a large-scale agent operating environment. The dashboard
plays a critical role in supporting operational oversight, performance evaluation, and continuous improvement, while
also informing decisions about when and how to safely increase the level of autonomy within agentic workflows.}

\textcolor{black}{Figure 6 illustrates the dashboard view, which presents a collection of analytic charts summarizing
agent execution patterns, lifecycle states, intervention frequencies, and HITL engagement metrics. These visualizations
allow users to quickly assess system health, identify bottlenecks, detect anomalous behavior, and understand where
human involvement is most frequently required. By consolidating this information at scale, the dashboard enables
organizations to manage thousands of concurrently operating agents in a controlled and informed manner.}

\textcolor{black}{In addition to static visualization, the dashboard integrates interactive analysis through a natural
language interface powered by a large language model. As shown in Figure 7, selecting a chart allows users to open an
LLM-driven chat window that generates a contextual analysis report explaining observed trends and patterns. Users can
further engage in dialogue with the LLM to ask follow-up questions, explore root causes, and derive business insights
related to efficiency, risk, and workflow optimization. This combination of visual analytics and conversational
analysis supports deeper understanding of agentic behavior and helps users identify targeted opportunities to refine
processes, improve safety, and incrementally advance the autonomy of the overall agentic workflow system.}

 \includegraphics[width=6.2681in,height=8.4957in]{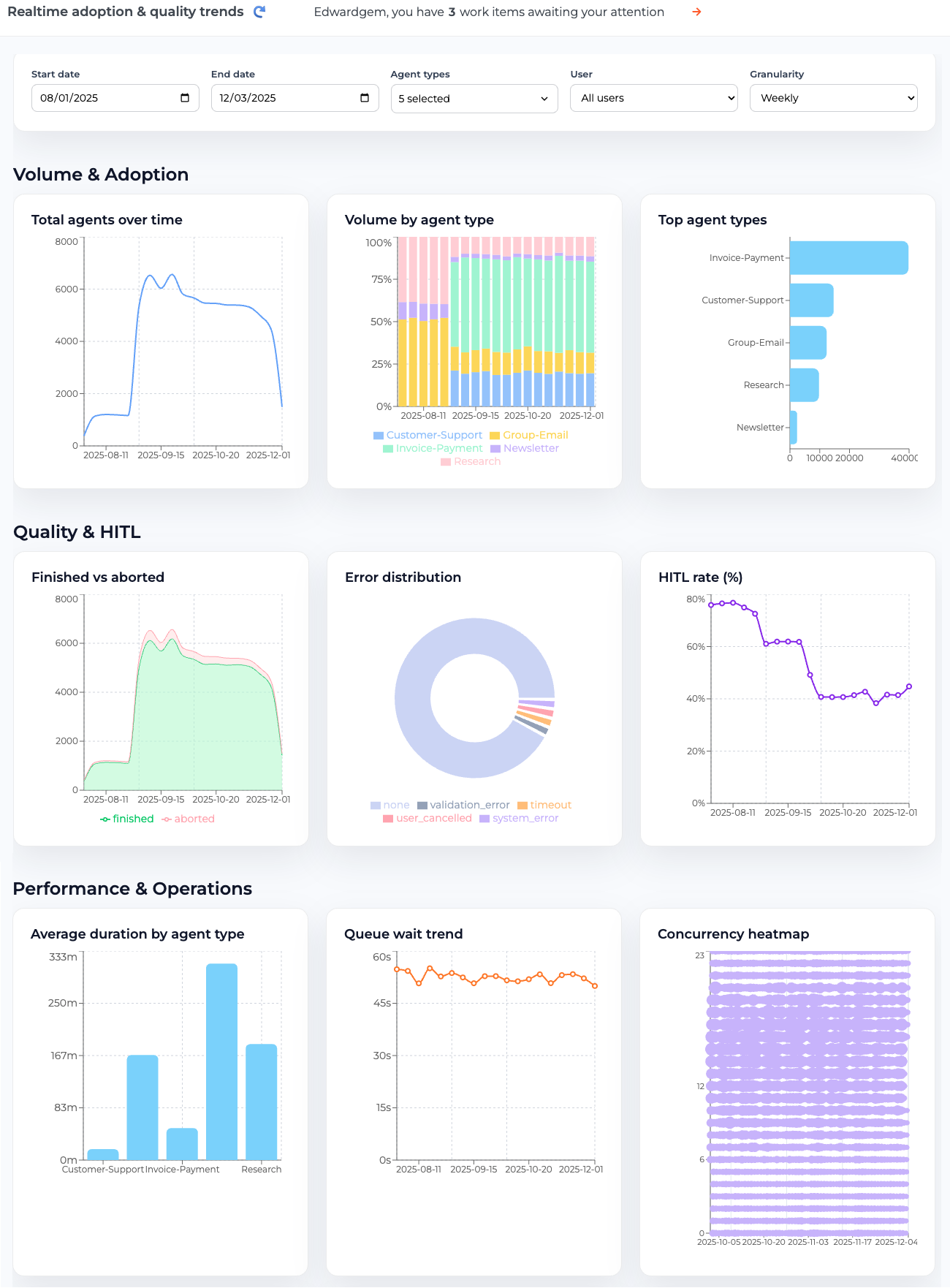} 

{\centering
\textbf{\textcolor{black}{Figure 6 Analytic Charts Illustrating Realtime Adoption and Quality Trends of the Agentic
System}}
\par}

\bigskip

{\centering  \includegraphics[width=2.8772in,height=6.5661in]{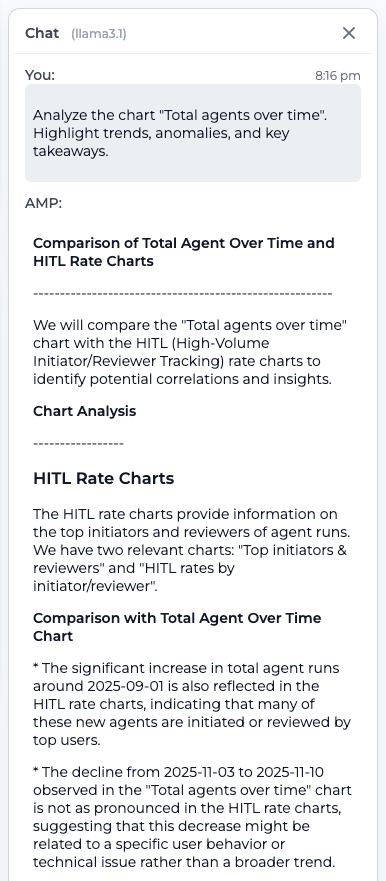} 
\includegraphics[width=2.7866in,height=6.5063in]{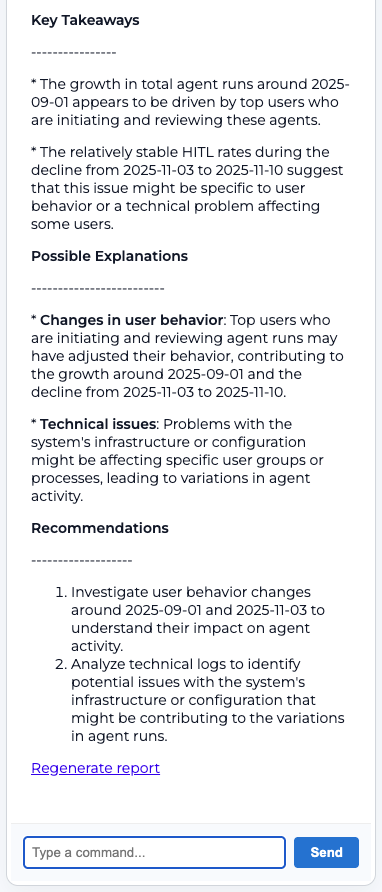} \par}
{\centering
\textbf{\textcolor{black}{Figure 7 LLM-Integrated Chat Interface Enabling Analytic Insights From Agent Dashboard
Charts}}
\par}

\liststyleWWNumiii
\setcounter{saveenum}{\value{enumi}}
\begin{enumerate}
\setcounter{enumi}{\value{saveenum}}
\item {\scshape\color{black}
Next Steps and Future Work}
\end{enumerate}
\textcolor{black}{The work presented in this paper establishes both a conceptual framework and an operational foundation
for developing safe, transparent, and trustworthy AI agents through the 3PM. Building on insights from prior research
in Human-in-the-Loop systems and safe AI, the model is extended into a comprehensive operational approach supported by
practical principles and implementation guidelines. As a practical and applicable framework, the 3PM is lightweight,
easy to understand, and straightforward to apply in the development, deployment, and operation of large-scale,
enterprise-grade agentic systems. At the same time, it is grounded in a complete and coherent theoretical foundation
and is designed to evolve as the scope, scale, and complexity of the agentic environment expand. The next phase of this
initiative focuses on translating these principles into real-world practice, ensuring that both industry and society
can fully benefit from the responsible adoption of autonomous agents. To achieve this, three primary work streams have
been initiated to extend, validate, and operationalize the ideas introduced in this study.}

\liststyleWWNumxi
\begin{enumerate}
\item {\bfseries\itshape\color{black}
Public Deliberation through the Stanford Deliberative Democracy Lab}
\end{enumerate}
\textcolor{black}{The first work stream involves a collaboration with the Deliberative Democracy Lab (DDL) at Stanford
University, which is conducting a series of public deliberative forums focused on the social and ethical dimensions of
AI agents [17, 18]. These forums bring together a diverse range of stakeholders, including AI industry leaders,
policymakers, researchers, and members of the public, to engage in structured, open discussions about the Three Pillars
of transparency, accountability, and trustworthiness.}

\textcolor{black}{In the initial phase, the DDL will conduct forums in North America and in India. The goal of this
initiative is to bridge the gap between technological innovation and societal readiness. By involving the public in
open, informed conversations, this work stream seeks to better understand how people perceive the risks and benefits of
autonomous agents, what level of transparency they expect, and what safeguards they require to build trust. The
insights from these dialogues will guide both technical and policy frameworks, ensuring that the development of AI
agents aligns with public values and expectations across both business and consumer contexts.}

\textcolor{black}{Through these deliberative processes, the AI community can establish mutual understanding and
legitimacy around agent governance, helping society evolve toward an era of AI-enabled collaboration rather than
resistance or fear.}

\liststyleWWNumxi
\setcounter{saveenum}{\value{enumi}}
\begin{enumerate}
\setcounter{enumi}{\value{saveenum}}
\item {\bfseries\itshape\color{black}
Industry Collaboration through the Safe AI Agent Consortium}
\end{enumerate}
\textcolor{black}{The second work stream focuses on industry collaboration through the Safe AI Agent Consortium, an
emerging alliance of leading organizations that share a commitment to advancing the responsible use of autonomous
agents [19]. The consortium's core members include Anthropic, Cohere, DoorDash, Meta, Microsoft, Oracle, PayPal,
Stanford, and other key players across academia, technology and enterprise sectors.}

\textcolor{black}{This group is jointly developing a set of industry guidelines and best practices grounded in the 3PM.
These guidelines aim to operationalize the concepts of transparency, accountability, and trustworthiness in a way that
developers, implementers, and users can readily apply to real-world AI systems. By creating common standards for agent
design, documentation, observability, and governance, the consortium seeks to promote safe adoption of AI agents at
scale. This initiative enables enterprises to capture productivity gains without compromising human oversight or public
trust.}

\textcolor{black}{The consortium's open work may also expand to developing shared benchmarks, safety testing protocols,
and interoperability frameworks for agent operating environments. These outcomes will serve as practical tools for both
startups and large organizations to evaluate the maturity, safety, and reliability of their agentic systems. Through
collective action and transparency among participants, this initiative aspires to make safety and responsibility a
competitive advantage in the growing agent economy.}

\liststyleWWNumxi
\setcounter{saveenum}{\value{enumi}}
\begin{enumerate}
\setcounter{enumi}{\value{saveenum}}
\item {\bfseries\itshape\color{black}
Open Tools and the Three-Pillar Agent Operating Environment}
\end{enumerate}
\textcolor{black}{The third work stream extends this research into applied development and community tooling. The
objective is for industry leaders and startups to design and release a set of open-source tools and frameworks that
embody the 3PM and accelerate the adoption of safe agentic systems. This includes the creation of an agent operating
environment, as illustrated in this paper, that integrates transparency, accountability, and trustworthiness by design
across the full agent lifecycle.}

\textcolor{black}{This environment will provide a standardized foundation for safe and effective agent operations,
offering key capabilities such as:}

\liststyleWWNumvii
\begin{itemize}
\item \textcolor{black}{Agent activity logging and lifecycle tracking to ensure full transparency and traceability
across initiation, execution and completion stages.}
\item \textcolor{black}{Decision journaling and explainability modules to support accountability by recording the
reasoning, context, and outcomes behind each agent decision.}
\item \textcolor{black}{Configurable human oversight controls and fallback mechanisms to maintain trustworthiness and
provide dynamic risk management through defined intervention thresholds.}
\item \textcolor{black}{AI generated analytics derived from agent activity logging and decision journals, with LLM
deployed throughout the 3PM operating environment to enable interactive monitoring, health assessment, and insight
generation. This capability allows users to better understand system behavior, identify improvement opportunities, and
make informed decisions about progressively increasing levels of agent autonomy.}
\item \textcolor{black}{AI-assisted 24x7 monitoring of agentic workflows to continuously learn behavioral patterns,
detect anomalies, and trigger timely human involvement when necessary to preserve system safety and security.}
\end{itemize}
\textcolor{black}{By providing a shared technical foundation, this work stream aims to lower the entry barrier for
organizations to adopt AI agents. It allows developers to embed safety and governance principles from the outset,
rather than retrofitting compliance and oversight after deployment. The tools will be open for collaboration and
extension by the research and developer communities, designed to integrate with existing agentic interoperability
standards such as the Model Context Protocol (MCP) and the Agent-to-Agent (A2A) communication protocol. This openness
will encourage cross-industry experimentation, validation, and interoperability, fostering a unified ecosystem where
safe, transparent, and accountable AI agents can evolve and operate seamlessly across different environments.}

\textcolor{black}{Through continuous contribution from the developer community and iterative improvement, the resulting
ecosystem will foster a trusted agent economy in which innovation can advance both responsibly and efficiently. Over
time, this environment may serve as a reference implementation for regulators, researchers, and practitioners seeking
to harmonize safety and governance standards across industries and geographic regions, thereby accelerating the safe
and scalable adoption of autonomous agents worldwide.}

\liststyleWWNumiii
\begin{enumerate}
\item {\scshape\color{black}
Conclusion}
\end{enumerate}

\bigskip

\textcolor{black}{This paper has presented a conceptual and operational framework for developing safe, transparent, and
trustworthy AI agents through the 3-Pillar Model (3PM), consisting of Transparency, Accountability, and
Trustworthiness. Building upon prior research in Human-in-the-Loop (HITL) systems, reinforcement learning with human
feedback, and collaborative AI, this model provides a practical foundation for guiding the evolution of AI agents from
assisted to fully autonomous operation. The framework emphasizes that autonomy must be achieved through a gradual,
verifiable process in which trust is earned over time, rather than assumed by design.}

\textcolor{black}{We have argued that the development of autonomous agents parallels the evolutionary path of autonomous
driving, where safety, reliability, and human confidence were cultivated through progressive stages of shared control.
Similarly, the journey toward trustworthy AI autonomy requires environments that support visibility, ethical reasoning,
and human collaboration. The proposed Three-Pillar Model ensures that every stage of agent development and deployment
remains transparent, accountable, and grounded in timely and appropriate human oversight. Transparency provides
observability into agent behavior and decision-making processes; accountability ensures that both actions and decisions
are traceable, explainable, and correctable; and trustworthiness transforms these safeguards into lasting confidence
among users, organizations, and the broader public.}

\textcolor{black}{To move from concept to practice, this research has initiated three complementary work streams. The
first engages the public through the Deliberative Democracy Lab at Stanford University, facilitating informed dialogue
between citizens and AI industry leaders about the social implications of agent transparency, accountability, and
trust. The second advances industry collaboration through the Safe AI Agent Consortium, uniting leading technology
organizations to establish shared best practices, evaluation benchmarks, and governance standards for safe agentic
systems. The third work stream focuses on open tooling, with the goal of developing an open agent operating environment
that embodies the Three-Pillar principles and supports interoperability among both native and external agents through
protocols, including the Model Context Protocol (MCP), Agent-to-Agent (A2A) communication, Agent Communication Protocol
(ACP), and Agent Network Protocol (ANP).}

\textcolor{black}{Through these efforts, the 3PM progresses from theoretical construct to actionable framework, enabling
the responsible evolution of autonomous agents. Through sustained collaboration across academia, industry, and society,
we can shape a future in which AI agents operate in alignment with human values, advancing innovation while upholding
safety, transparency, and ethical integrity.}

\bigskip

{\centering
\textbf{\textsc{\textcolor{black}{References}}}
\par}

\liststyleWWNumix
\begin{enumerate}
\item \textcolor{black}{Sanders, T. }\textit{\textcolor{black}{How AI Agents Are Overcoming Market Hype to Deliver Real
Business Impact.}}\textcolor{black}{ 2025 AI Agents G2 Insight Report, October 2025.
}\url{https://company.g2.com/news/2025-ai-agent-report}\textcolor{black}{ }
\item \textcolor{black}{Zanzotto, F.M. }\textit{\textcolor{black}{Viewpoint: Human-in-the-loop Artificial
Intelligence.}}\textcolor{black}{ Journal of Artificial Intelligence Research 64 (2019) 243-252. February 2029.}
\item \textcolor{black}{Wu, X. et al. }\textit{\textcolor{black}{A Survey of Human-in-the-Loop for Machine
Learning.}}\textcolor{black}{ arXiv:2108.00941 (v3). April 2022.
}\url{https://arxiv.org/abs/2108.00941}\textcolor{black}{ }
\item \textcolor{black}{Mosqueira\nobreakdash-Rey, E. et al.
}\textit{\textcolor{black}{Human\nobreakdash-in\nobreakdash-the\nobreakdash-Loop Machine Learning: A State of the
Art.}}\textcolor{black}{ Artificial Intelligence Review (2023) 56:3005--3054. August 2022.
}\url{https://link.springer.com/article/10.1007/s10462-022-10246-w}\textcolor{black}{ }
\item \textcolor{black}{Wixom, B., Someh, I., and Gregory, R. }\textit{\textcolor{black}{AI Alignment: A New Management
Paradigm.}}\textcolor{black}{ MIT Center for Information Systems Research (MIT CISR). No. XX-11. November 2020.
}\url{https://cisr.mit.edu/publication/2020_1101_AI-Alignment_WixomSomehGregory}\textcolor{black}{ }
\item \textcolor{black}{Burnham, K. }\textit{\textcolor{black}{New framework helps companies build secure AI
systems.}}\textcolor{black}{ MIT Management Sloan School. July 2025.
}\url{https://mitsloan.mit.edu/ideas-made-to-matter/new-framework-helps-companies-build-secure-ai-systems}\textcolor{black}{
}
\item \textcolor{black}{Bellos, F. et al. }\textit{\textcolor{black}{Towards Effective Human-in-the-Loop Assistive AI
Agents.}}\textcolor{black}{ arXiv:2507.18374 (v1). July 2025. }\url{https://arxiv.org/abs/2507.18374}\textcolor{black}{
}
\item \textcolor{black}{Mozannar, H. et al. }\textit{\textcolor{black}{Magentic-UI: Towards Human-in-the-loop Agentic
Systems.}}\textcolor{black}{ Microsoft Research AI Frontiers. arXiv:2507.22358 (v1). July 2025.
}\url{https://arxiv.org/abs/2507.22358}\textcolor{black}{ }
\item \textcolor{black}{Retzlaff, C. O. et al. }\textit{\textcolor{black}{Human-in-the-Loop Reinforcement Learning: A
Survey and Position on Requirements, Challenges, and Opportunities. }}\textcolor{black}{Journal of Artificial
Intelligence Research 79 (2024) 359-415. January 2024.}
\item \textcolor{black}{Ren, A. Z. et al. }\textit{\textcolor{black}{Robots That Ask For Help: Uncertainty Alignment for
Large Language Model Planners.}}\textcolor{black}{ Google DeepMind. arXiv:2307.01928 (v2). September 2023.
}\url{https://arxiv.org/abs/2307.01928}\textcolor{black}{ }
\item \textcolor{black}{Allen, D. et al. }\textit{\textcolor{black}{A Roadmap for Governing AI: Technology Governance
and Power Sharing Liberalism.}}\textcolor{black}{ ASH Center for Democratic Governance and Innovation. Harvard Kennedy
School. January 2024.
}\url{https://ash.harvard.edu/wp-content/uploads/2024/01/340040_hks_ashgovroadmap_v2.pdf}\textcolor{black}{ }
\item \textcolor{black}{Barroso, L. R. and Mello, P. P. C. }\textit{\textcolor{black}{Artificial Intelligence: Promises,
Risks, and Regulation: Something New Under the Sun.}}\textcolor{black}{ CARR Center for Human Rights Policy. Harvard
Kennedy School. December 2024.
}\url{https://www.hks.harvard.edu/sites/default/files/2024-12/24_Barroso_Digital_v3.pdf}\textcolor{black}{ }
\item \textcolor{black}{Natarajan, S. et al. }\textit{\textcolor{black}{Human-in-the-loop or AI-in-the-loop? Automate or
Collaborate?}}\textcolor{black}{ The Thirty-Ninth AAAI Conference on Artificial Intelligence (AAAI-25). March 2025.}
\item \textcolor{black}{Wang, J.,~Zhang, L., Huang, Y.,~and Zhao, J. }\textit{\textcolor{black}{Safety of Autonomous
Vehicles. Journal of Advanced Transportation. }}\textcolor{black}{October 2020.
}\url{https://doi.org/10.1155/2020/8867757}\textcolor{black}{ }
\item \textcolor{black}{Khan, M. A. et al. }\textit{\textcolor{black}{Level-5 Autonomous Driving---Are We There Yet?
}}\textcolor{black}{A Review of Research Literature. ACM Journals, ACM Computing Surveys (CSUR), Vol. 55, Issue 2,
Article No. 27. January 2022. }\url{https://doi.org/10.1145/3485767}\textcolor{black}{ }
\item \textcolor{black}{Cheng, J. }\textit{\textcolor{black}{Context-Aware Prompt Enhancement (CAPE) Framework for a
Multi-Agent Application System.}}\textcolor{black}{ Inquiryon, Inc. July 2025.}
\item \textcolor{black}{Siu, A}\textit{\textcolor{black}{. Industry-Wide Deliberative Forum Invites Public to Weigh In
on the Future of AI Agents.}}\textcolor{black}{ First public announcement. June 2025.
}\url{https://deliberation.stanford.edu/industry-wide-deliberative-forum-invites-public-weigh-future-ai-agents}\textcolor{black}{
}
\item \textcolor{black}{Siu, A. }\textit{\textcolor{black}{DoorDash and Microsoft join Industry-Wide Deliberative Forum
on Future of AI Agents.}}\textcolor{black}{ Second public announcement. August 2025.
}\url{https://deliberation.stanford.edu/doordash-and-microsoft-join-industry-wide-deliberative-forum-future-ai-agents}\textcolor{black}{
}
\item \textcolor{black}{Katsanevas, A. et al. }\textit{\textcolor{black}{AI Agent for Good: Alignment, Safety, \&
Impact.}}\textcolor{black}{ 2025 Summer Symposium Hosted by Stanford Deliberative Democracy Lab. July 2025.
}\url{https://deliberation.stanford.edu/ai-agent-good-alignment-safety-impact}\textcolor{black}{ }
\end{enumerate}
\end{document}